\documentclass[twocolumn,amssymb,nobibnotes,pra,superscriptaddress]{revtex4}
  \usepackage{color}
  \usepackage{newlfont}
  \usepackage{graphicx}
  \usepackage{amssymb}
  \usepackage{amsmath}
  \usepackage[latin1]{inputenc}
  \usepackage{float}
 \usepackage{mathpazo}

\newcommand{\ket}[1]{\mbox{$ | #1 \rangle $}}
\newcommand{\bra}[1]{\mbox{$ \langle #1 | $}}

\newcommand{\cm}{{cm}}
\newcommand{\rr}{{r}}
\newcommand{\ri}{{i}}
\newcommand{\rp}{{p}}

\renewcommand{\r}{{\textrm{r}}}		
\newcommand{\ba}{\begin{eqnarray}}
\newcommand{\ea}{\end{eqnarray}}
\newcommand{\proj}[1]{\ket{#1}\!\bra{#1}}
\newcommand{\op}[2]{| #1 \rangle \! \langle #2 |}
\newcommand{\tr}{{\textrm{tr}}}

\begin{document}

\title{Physics within a quantum reference frame}

\author{Renato M. Angelo}
\affiliation{Federal University of Parana, P.O. Box 19044, 81531-990, Curitiba, PR, Brazil.}
\affiliation{H.H. Wills Physics Laboratory, University of Bristol, Tyndall Avenue, Bristol, BS8 1TL, United Kingdom}
\author{Nicolas Brunner}
\affiliation{H.H. Wills Physics Laboratory, University of Bristol, Tyndall Avenue, Bristol, BS8 1TL, United Kingdom}
\author{Sandu Popescu}
\affiliation{H.H. Wills Physics Laboratory, University of Bristol, Tyndall Avenue, Bristol, BS8 1TL, United Kingdom}
\author{Anthony J. Short}
\affiliation{DAMTP, Centre for Mathematical Sciences, Wilberforce Road, Cambridge, CB3 0WA, United Kingdom}
\author{Paul Skrzypczyk}
\affiliation{H.H. Wills Physics Laboratory, University of Bristol, Tyndall Avenue, Bristol, BS8 1TL, United Kingdom}

\begin{abstract}
We investigate the physics of quantum reference frames. Specifically, we study several simple scenarios involving a small number of quantum particles, whereby we promote one of these particles to the role of a quantum observer and ask what is the description of the rest of the system, as seen by this observer? We highlight the interesting aspects of such questions by presenting a number of apparent paradoxes. By unravelling these paradoxes we get a better understanding of the physics of quantum reference frames.
\end{abstract}

\maketitle

\section{Introduction}

Reference frames are one of the most basic notions in physics. Almost every time a measurement is performed or a formula is written down, a reference frame has been implicitly used. The choice of  reference frame can either be thought of as abstract labelling of space-time, or a description relative to some physical laboratory with a particular set of rulers and clocks. In the latter case, the laboratory is  always taken to be much heavier than the system under observation, with a well defined velocity and position. In other words, it is assumed to be fundamentally {\em classical}.

However, in their seminal papers \cite{Suss67a,Suss67b}, Aharonov and Susskind showed that the concept of reference frame can be suitably accommodated in quantum theory. Although our everyday reference frames are classical for all practical purposes, Aharonov and Susskind have shown that theoretically nothing prevents the idea of a {\em quantum reference frame}. These early works were followed by \cite{carmi73,carmi74} and culminated in the work of Aharonov and Kaufherr \cite{aharonov84} that is of particular relevance to us here.

	Since then much progress has been made by investigating the role of quantum reference frames from both foundational and applied viewpoints. Several works have investigated fundamental properties of quantum reference frames \cite{poulin07,BSLB08,GP08,GS08,GMS09,Rez09}. Also, quantum reference frames have been studied in quantum information \cite{bartlett07}, in particular in the context of quantum communication \cite{Massar95,Gisin99,PeresScudo01,Bagan00, bartlett09,ioannou09} and entanglement detection \cite{costa09,liang09}.

In the present paper we go back to the original problem discussed in \cite{aharonov84}, namely that of relative positions and momenta, and ask what physics looks like from an isolated quantum reference frame. In particular, we  consider states composed of a small number of quantum systems, then promote one of these systems to the role of a \emph{quantum observer} and examine the description of the state from its internal perspective. For example, we will consider experiments performed within a freely-floating quantum rocket, from the perspective of the rocket itself , making no mention of any external reference frame whatsoever. We find that we are easily lead into paradoxes when trying to analyse  experiments from the perspective of such a quantum observer. By careful reanalysis, we resolve these paradoxes and in doing so learn much about the physics of quantum reference frames.

\section{Quantum observers in superposition}
\label{sec:2_particle}

\begin{figure}[b]
        \includegraphics[width=\columnwidth]{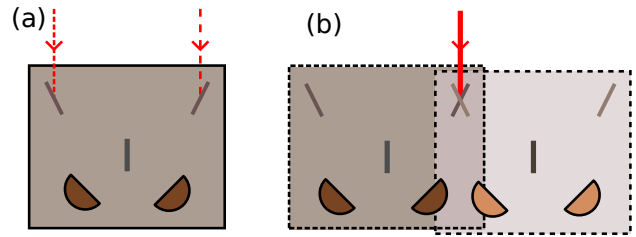}
    \caption{Interference experiment in which (a) a particle is sent in superposition towards an interferometer; (b) a particle is sent towards an interferometer which is in superposition.}
    \label{f:mz}
\end{figure}

Let us consider a simple interference experiment, in which a
particle is sent in superposition towards an interferometer.
Although such experiments usually involve photons, for simplicity
here we consider the particle to be non-relativistic. The
interferometer consists of two mirrors, a beam splitter, and two
detectors, as depicted in Fig.~\ref{f:mz} (a). We take these
components to be rigidly fixed together, but free to move relative
to the laboratory (which we treat as a classical external reference
frame).

We initially place the centre of the interferometer at the origin of
the laboratory, and denote the position of the left and right path
of the particle by $-L$ and $L$ respectively. The initial state of
the interferometer and particle relative to the laboratory can then be written
\begin{equation}
\ket{\psi_a} =
\ket{0}_{i}\left(\frac{\ket{-L}_{p}+\ket{L}_{p}}{\sqrt{2}}\right),\label{eqn:setup_a}
\end{equation}
where $\ket{x}_{i/p}$ denotes a narrow wavepacket for the
interferometer/particle centred at $x$ (in the horizontal coordinate).

This description of the experiment is from the point of view of the
laboratory, but what we are actually interested in is the
description of this experiment from the perspective of the
interferometer and the particle themselves. Since neither the
interferometer nor the particle interacts with the laboratory we
should not need to use it to describe the experiment. To this end we
ask the following questions: a) What does the interferometer see? b)
What does the particle see?

Let us begin by first clarifying precisely what we mean by our terminology that the the interferometer or particle `sees'. We use this rather colloquial terminology to mean the description of the physics relative to the system in question, i.e. the relative positions and momenta of the rest of the system. 

Assuming that the interferometer is much heavier than the particle,
the answers are as we might expect. The interferometer sees the
particle approaching it in a superposition of positions $-L$ and $L$,
while the particle sees the inverse -- the interferometer approaching
it in a superposition of positions $L$ and $-L$.

Now consider the slightly different experimental setup shown in
Fig.~\ref{f:mz} (b). In this case, the particle is positioned at the
origin of the laboratory, while the centre of the interferometer is
placed in a superposition of positions $-L$ and $L$. The state of
the joint system for this second experiment is
\begin{equation} \label{eqn:setup_b}
\ket{\psi_b} =
\left(\frac{\ket{-L}_{i}+\ket{L}_{i}}{\sqrt{2}}\right)\ket{0}_{p}.
\end{equation}

One might think that as far as the particle and interferometer are
concerned this second situation would look the same as the first
experiment, with the interferometer seeing the particle approaching
it in a superposition of positions $-L$ and $L$, and the particle
seeing the interferometer in a superposition of $L$ and $-L$ as
before. Since it is only the particle and interferometer that are relevant, and not the external frame, we would expect the two experiments to yield
identical results.

Indeed, if we were to place the detectors instead at the location of the mirrors then both situations would be equivalent; in both cases both detectors would be equally likely to click.

Surprisingly, however, if we follow the evolution in both cases we see a dramatic difference. This is illustrated from the
perspective of the laboratory in figure 2. In the first experiment, the two particle paths interfere at the beam splitter and only
the left detector clicks, while in the second experiment there is no interference between the photon paths and both detectors are
equally likely to click.

\begin{figure}[h]
        \includegraphics[width=\columnwidth]{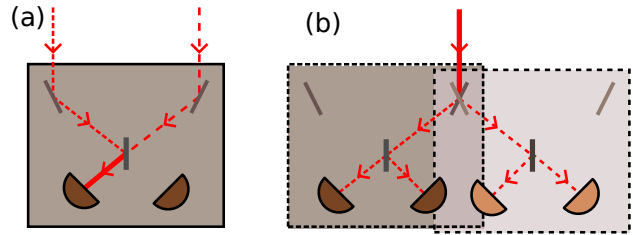}
    \caption{In (a) the photon, initially in superposition, interferes constructively at the beam splitter and always exits through the left output port and only the left detector ever clicks. In (b) it is the interferometer which is in superposition. In each branch of the superposition the photon enters the beam splitter through a single input port, hence no interference occurs and both detectors are equally likely to click.}
    \label{f:mz2}
\end{figure}

As all perspectives must agree on the probability of detector
clicks, the second experiment must also look different to the first
from the perspective of the particle and the interferometer. In other words, experiments (a) and (b) are not equivalent, contrary to what we expected. But why not?

To find out, we consider in more detail the correct description of
the particle relative to the interferometer for both experiments. To
do this we change from laboratory co-ordinates to the relative
coordinate of the particle as seen by the interferometer, $ x_r $, and the centre of mass, $x_{\cm}$. These are given by

\begin{eqnarray}
    x_{\cm} &=& \frac{m_{i}x_{i}+m_{p}x_{p}}{M} \label{eqn:x_cm} \\ x_r &=&
    x_{p}-x_{i}, \label{eqn:x_r}
\end{eqnarray}
where $m_{i/p}$ and $x_{i/p}$ are the mass and position of
the interferometer/particle, and $M = m_i + m_p$.

One should note that when changing co-ordinates subtle effects due to the finite width of the wavepackets emerge. Consider a state of the form $\ket{a}_1 \ket{b}_2$, by which we actually mean $\psi(x_1-a)\phi(x_2-b)$, i.e. two wavepackets, centred around $a$ and $b$, hence the centre of mass is located around $\alpha = (m_1 a + m_2 b)/M$ and the relative co-ordinate is centred around $\beta = b-a$. However, in this state there is an uncertainty in the position of each particle, hence there is an uncertainty in the position of the centre of mass and relative distance; in fact the centre of mass and relative position will in general be entangled. More precisely, in relative co-ordinates  the state $\psi(x_1-a)\phi(x_2-b)$ becomes  $\Psi(x_\cm, x_\rr) =  \psi(x_\cm+m_2 x_\r /M - a)\phi(x_\cm-m_1 x_\r/M -b)$ which will in general not factorise as there will be entanglement between the \emph{fluctuations} of the centre of mass and relative coordinate. Entanglement between the centre of mass and relative distance will turn out to be crucial in what follows. However, it is not this entanglement between the fluctuations but the entanglement that appears when we consider superpositions of localised wavepackets, such as $\ket{a}_1 \ket{b}_2 + \ket{a'}_1 \ket{b'}_2$ that is crucial in our paper. In other words, our concern is the entanglement between the \emph{mean} positions of the centre of mass and relative co-ordinates, not the entanglement of the fluctuations. For the sake of clarity, from here on we thus work under the following approximation, $\Psi(x_\cm,x_\rr) \simeq \psi'(x_\cm-\alpha)\phi'(x_\rr-\beta) = \ket{\alpha}_{\cm}\ket{\beta}_\rr$, which amounts to ignoring the entanglement between the fluctuations. Importantly, this approximation has no consequences on the physics we present below; for a more detailed analysis, including finite width effects, see the appendix.

Given the above, in  relative co-ordinates the two experimental situations are described by the states
\begin{eqnarray}
    \ket{\psi_a} &\simeq& \frac{\ket{\frac{-m_p L}{M}}_{cm} \ket{-L}_r+\ket{\frac{m_p L}{M}}_{cm}\ket{L}_r}{\sqrt{2}} \\
    \ket{\psi_b} &\simeq& \frac{\ket{\frac{-m_i L}{M}}_{cm} \ket{L}_r+\ket{\frac{m_i L}{M}}_{cm}\ket{-L}_r}{\sqrt{2}}.
\end{eqnarray}
The crucial point is to realise that there are degrees of freedom that are accessible to the internal observer and degrees of freedom which are not and to distinguish between them. On the one hand, it is clear
that the position of the centre of mass is a degree of freedom that
is not accessible inside  as it explicitly refers to the external reference frame, i.e. the laboratory. Indeed the position of the centre
of mass can neither be determined nor altered by any internal
observer that has no access to any external frame. On the other hand, the relative position between the particle and interferometer is certainly accessible as it does not refer to the laboratory (or any other external frame).
Therefore to correctly describe the physics of the internal observer we must trace over the inaccessible degrees of freedom, i.e.
the centre of mass. It is the reduced state, $\rho_p =
\textrm{Tr}_{cm}(\ket{\psi}\bra{\psi})$ which accurately describes
the state of the particle relative to the interferometer. Similarly
by taking instead $x_r=x_i-x_p$ we find the state of the
interferometer relative to the particle.

Now, if we consider the regime where the mass of the interferometer
is much greater than that of the particle, that is $m_\rp \ll m_\ri$
\footnote{In particular, if the centre of mass wavepacket is of size
$\epsilon$, we require $\frac{m_p}{m_i} \ll
\frac{\epsilon}{|L|}$} then we see that the states (5) and (6)
become, approximately,
\begin{align}
    \ket{\psi_a}&\simeq \ket{0}_{cm}\left(\frac{\ket{-L}_r+\ket{L}_r}{\sqrt{2}}\right) \\
    \ket{\psi_b}&\simeq \frac{\ket{-L}_{cm}\ket{L}_r+\ket{L}_{cm}\ket{-L}_r}{\sqrt{2}}
\end{align}
Thus we see that in experiment (a) upon tracing over the centre of
mass we are left in the pure state
$\frac{\ket{-L}_r+\ket{L}_r}{\sqrt{2}}$. In this
situation the interferometer sees the particle in a coherent
superposition, hence we observe interference and only one of the
detectors clicks. On the other hand however, in experiment (b) upon
tracing over the centre of mass we are left in a mixed state
$\frac{\ket{L}_{r}\bra{L} + \ket{-L}_r\bra{-L}}{2}$ representing an equal
chance of the particle impinging upon the left and right mirrors.
Here no interference can take place and hence both detectors are
equally likely to click.

Note that the relative state of the interferometer as seen by the
particle is exactly the same, up to the sign of the relative
position, and thus the same conclusions also hold from this
perspective. Hence both internal observers will make the same
predictions as the external observer about the experimental outcome.
We see that the reason why the internal observers make different
predictions in the two experiments is due to the fact that relative
states in each experiment are different; one is a coherent
superposition, whilst the other is a mixture.

We are thus led to realise that the centre of mass plays a crucial
role when considering what the interferometer or particle sees, for
it is the entanglement of the relative coordinate with the centre of
mass coordinate which makes the two states (5) and (6) completely
different.

Below, we provide another thought-experiment which shows explicitly how interference disappears whenever the centre of mass is entangled with the internal degrees of freedom. This is in order to make it clear that the disappearance of interference is due to the entanglement between the centre of mass and relative co-ordinate which was already present in the initial state of the system, and not due to its subsequent interactions, such as the reflection of the particle by the mirrors.

\subsection{Inside a quantum rocket} \label{sec:rocket}

\begin{figure}[ht]
\includegraphics[width=\columnwidth]{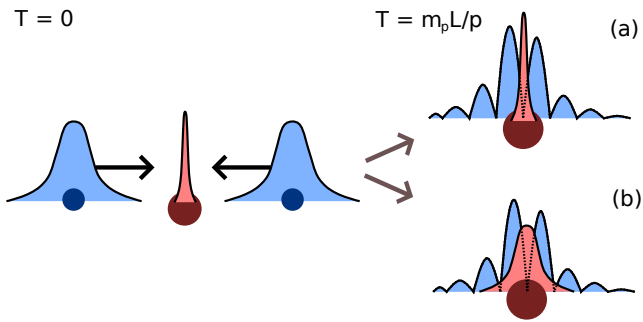}
\caption{At $T=0$ a quantum rocket is prepared in a well localised state, whose centre of mass is here depicted by the narrow red wavepacket. Inside the rocket is a particle in superposition, here depicted as the blue wavepackets, initially located a distance $L$ from the centre of the rocket, with momemtum $p$ towards the centre. At a time $T = m_p L/p$ when the wavepackets interfere either (a) the centre of mass is sufficienty localised to resolve the fringes or (b) the centre of mass is too uncertain to resolve them. \label{rocket}}
\end{figure}

Consider the situation depicted in Fig.~\ref{rocket} in which a particle of mass $m_p$ moves freely inside a stationary rocket of mass $m_R \gg m_p$. After the preparation of their initial state, the particle and the rocket no longer interact with the external world. For an external observer this is a {\em quantum} rocket, i.e. the position of the rocket's centre of mass is given by a narrow wavepacket centred at the origin of the external frame, with a quantum uncertainty $\Delta x_R$.

Suppose the particle has been prepared in a coherent superposition relative to the external observer. The two wave packets composing the superposition move towards each other with constant average momentum $p$. Initially, they are symmetrically located at a distance $L$ apart from the origin of the external frame.

After a period of time $T$, given by
\begin{eqnarray}\label{T}
T=m_p L/p,
\end{eqnarray}
the two wave packets meet at the origin and from the external observer's perspective, they form an interference pattern corresponding to the  wavelength $\lambda=2\pi/p$, in units such that $\hbar=1$. Then the question arises whether an observer inside the rocket would be able to see the interference pattern.

Intuitively we expect the uncertainty in the rocket's position at the time the two wavepackets interfere to be the limiting factor on the ability of the internal observer to measure the interference, that is the rocket can only resolve details significantly larger than its uncertainty $\Delta x_R(T)$ relative to the external frame. Thus, this implies that the internal observer can only see the interference pattern when the uncertainty in the rocket's position is much smaller than the wavelength $\lambda$ of the particle, namely
\begin{eqnarray}\label{condition}
\Delta x_R(T)\ll \lambda,
\end{eqnarray}
where $\Delta x_R(T)$ denotes the variance in the position of the rocket at the instant $T$. This is situation (a) in FIG. \ref{rocket}. On the other hand, in situation (b) when $\Delta x_R(T)\gg \lambda$, although from the external frame of reference there is still interference, from inside the rocket this can no longer be seen. What we shall now see is that whether we end up in situation (a) or (b) depends only upon whether or not the centre of mass of the rocket is entangled with the relative co-ordinates.

Let us assume that the rocket is initially in a minimum uncertainty wave packet ($\Delta x_R\Delta p_R=\frac{1}{2}$). Using the approximation $\Delta x_R(T)\simeq \Delta x_R+\Delta p_R T/m_R$ we now re-write Eq.~\eqref{condition} as
\begin{eqnarray}
\Delta x_R+\frac{T}{2m_R\Delta x_R}\ll \frac{2\pi}{p}.
\end{eqnarray}
Note that both terms on the left-hand side are positive, hence each must be much smaller than the right-hand side.  Using equation \eqref{T} this implies that
\begin{eqnarray}\label{condition2}
\frac{m_p L}{m_R}\ll \Delta x_R \ll \lambda.
\end{eqnarray}
This means that the interference pattern can indeed be resolved by an observer in the rocket provided that the initial uncertainty in the rocket's position obeys these relations.

Given that we take the mass of the rocket to be much larger than the particle, then $M \approx m_R$ and $\Delta x_{cm} \approx \Delta x_R$, and Eq. \eqref{condition2} implies that
\begin{equation}
	\frac{m_p L}{M} \ll \Delta x_{\cm}
\end{equation}
Here $\pm m_p L/M$ represents by how much the centre of mass is displaced from the origin when the particle is prepared to the left or to the right. Therefore the meaning of (13) is that this displacement is smaller than the original uncertainty in the centre of mass, hence the centre of mass is insensitive to whether the particle is to the left or the right, i.e. it decouples from the relative position.

Again we see that the entanglement between the centre of mass and the relative co-ordinates determines whether or not there will be interference. However, this is by no means the end of the story, as we shall see in the next section.

\section{Paradox of the third particle}

In the previous section we have seen that the centre of mass, in particular its entanglement to the relative co-ordinates, plays a central role in the physics of quantum reference frames. However, we shall see now that this is not the only issue in understanding quantum reference frames. Contrary to the situation considered so far we shall envisage a thought experiment in which the centre of mass is physically irrelevant, yet we run into an even more intriguing paradox.

Let us consider now two
particles, of mass $m_1$ and $m_2$, prepared in the following
entangled state:
\begin{eqnarray} \label{phi}
\ket{\psi} =\frac{1}{\sqrt{2}}\Big(\ket{-a}_{1}\ket{b}_{2}+e^{i
\theta}\ket{a}_{1}\ket{-b}_{2} \Big)
\end{eqnarray}
where the state $\ket{x}_{i}$ denotes that particle $i$ is sharply
localized at position $x$ relative to the laboratory, and $\theta$
is an arbitrary relative phase.

\begin{figure}[h!]
    \includegraphics[width=\columnwidth]{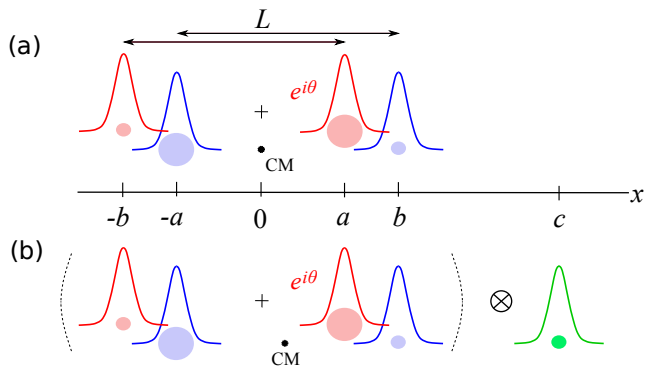} \caption{\label{f:particles} (a) The two particles are in an entangled state, such that the centre of mass of the system is unentangled from the relative co-ordinates. (b) A third, rather innocuous, particle is added in the picture, leaving the centre of mass unentangled with the relative co-ordinates.}
\end{figure}

We choose the positions $a$ and $b$ such that $m_1a = m_2b$ and
hence the centre of mass of the particles is at the origin of our
co-ordinate system, as shown in Fig.~\ref{f:particles}a. We also
denote the distance between the particles by $L = a+b$, which allows
us to write
\begin{equation}\label{ab}
a=\frac{m_2}{m_1+m_2}L, \qquad b=\frac{m_1}{m_1+m_2}L.
\end{equation}

So far the description of the system has been given relative to an
external reference frame. Again we would like to promote particle 1 to the role
of a \emph{quantum observer} and understand how \emph{it} would
describe the state of particle 2.

Thus we move again from absolute to relative positions. We therefore describe our system using the position of the centre
of mass $x_{cm}$ and the relative position $x_{r_2}$, given by
\begin{equation}
    x_{cm} = \frac{m_1x_1 + m_2x_2}{m_1+m_2} \qquad x_{r_2} = x_2-x_1.
\end{equation}
We can rewrite
the state of the system \eqref{phi} in these new co-ordinates as
\begin{equation}\label{Phi}
\ket{\psi} \simeq \ket{0}_{cm}\left(\frac{\ket{L}_{r_2}+e^{i
\theta}\ket{-L}_{r_2}}{\sqrt{2}}\right).
\end{equation}
Now we are in position to describe what particle 1 would observe
without access to the external reference frame, i.e. by tracing over the centre of mass. First, from Eq.
\eqref{Phi} we see that the position of the centre of mass and the
relative position actually decouple, i.e. the global state is a
product state. Therefore we conclude that particle 1 sees particle 2
in a pure state. Importantly this implies that particle 1 can get
access to the phase $\theta$ by interacting with particle 2 alone,
i.e. without access to the external reference frame.

Now let us bring a third, and rather innocuous, particle into the
game. This particle has mass $m_3$, and is located at position $c$
relative to the external frame (see Fig.~1b). Particles 1 and 2 are
exactly as previously. In the external reference frame, the global
state of the system is now given by
\begin{equation}\label{psi}
\ket{\Psi}=\frac{1}{\sqrt{2}}\Big(\ket{-a}_1\ket{b}_2+e^{i
\theta}\ket{a}_1\ket{-b}_2 \Big)\ket{c}_3
\end{equation}
In order to understand what particle 1 would observe, we move again
to relative co-ordinates, given by
\begin{eqnarray} \label{qj} \nonumber  x_{cm} &=&  \frac{m_1x_1 + m_2x_2+m_3x_3}{M} \\
    x_{r_2} &=& x_2-x_1 \\\nonumber
    x_{r_3} &=& x_3-x_1 \end{eqnarray}
where $M=m_1+m_2+m_3$. Note that $x_{r_2}$ and $x_{r_3}$ are the
positions of particle 2 and 3 relative to particle 1. The state of
the system, expressed in these new co-ordinates is
\begin{eqnarray} \label{Psi}
\ket{\Psi}\simeq\left|\frac{m_3}{M}c\right\rangle_{cm}\left(\frac{\ket{L}_{r_2}\ket{c+a}_{r_3}+e^{i
\theta}\ket{-L}_{r_2}\ket{c-a}_{r_3}}{\sqrt{2}} \right).
\end{eqnarray}

Again, let us distinguish those degrees of freedom which are
accessible to particle 1 from those which are not. First of all, we
note that the inaccessible degree of freedom, i.e. the position of
the centre of mass, factorizes once again. However, the surprising fact now is that the accessible degrees of
freedom---the positions of particles 2 and 3 relative to particle 1
---  are now entangled. Consequently it appears that particle 1 must
do a joint measurement on particles 2 and 3 to determine the phase
$\theta$. If particle 1 cannot access particle 3, for instance when
these two particles are far apart, it seems that we should trace
over $x_{cm}$ and $x_{r_3}$, giving the reduced state
\begin{equation} \label{red_state}
\rho_{r_2} = \mathrm{Tr}_{cm, r_3} (\ket{\Psi}\bra{\Psi})\simeq
\frac{\ket{L}\bra{L} + \ket{-L}\bra{-L}}2,
\end{equation}
which contains no information about the phase whatsoever.

Here the paradox emerges. Since the innocuous particle 3 is
completely uncorrelated from particles 1 and 2, adding it into the
description of the system should clearly not change the physics.
However from the above discussion, it appears that the addition of
particle 3 leads to dramatically different situations from the
perspective of the internal observer. On the one hand, when
considering only particles 1 and 2, we concluded that particle 1 can
gain information about the phase $\theta$. On the other hand, when
particle 3 is included (but inaccessible to particle 1), it appears
that particle 1 cannot extract any information about the phase
$\theta$.

Clearly this is in flagrant conflict with the predictions of the
external observer, who would argue that particle 1 should be able to
extract exactly the same information about $\theta$ in both cases,
as it is a property of particles 1 and 2 alone. In the following
subsections we shall identify the optical illusion which is actually present and hence solve the
paradox. In doing so we will learn more about the nature of quantum
observers.

\subsection{Measuring the phase}

One may suspect that the root of the problem resides in the phase measurement; indeed above we never explicitly showed how the phase could be measured. In the following we address this aspect in detail.

We focus initially on the case of two particles, for which we
concluded that particle 1 has access to the phase, and see how
particle 1 would proceed. Let us consider again the state of the
system in relative co-ordinates, i.e. Eq. \eqref{Phi}. Measuring the
relative position of particle 2 provides no information about the
phase. Instead, the information about the phase is encapsulated in the relative momentum,
\begin{equation}
\hat{p}_{r_2}=\mu_{12}\left(\frac{\hat{p}_2}{m_2}-\frac{\hat{p}_1}{m_1}\right),
\end{equation}
where $\mu_{12}=\frac{m_1m_2}{m_1+m_2}$ is the reduced mass. However, this information is encoded in an intricate way: The expectation values of the moments of momentum, $\langle \hat{p}_{r_2}^n\rangle$, are all independent of the relative phase. The function of momentum whose expectation is sensitive to the phase is the relative shift operator
$e^{-i 2L\hat{p}_{r_2}}$ where $\hbar = 1$ \cite{aharonov05,aharonov69}.

 This operator has
the effect of shifting the position of particle 2 versus particle 1
by a distance $2L$, causing one wavepacket in the original state to
overlap the other in the shifted state, and revealing information
about $\theta$ in its expectation value:
\begin{eqnarray}
	\label{phase}\nonumber \bra{\psi} e^{-i 2L\hat{p}_{r_2} } \ket{\psi}&\simeq& \tfrac{1}{2}(_{r_2}\bra{L}+e^{-i\theta}
	_{\quad\:r_2}\bra{-L}) \times \\
&& e^{-i 2L\hat{p}_{r_2}}(\ket{L}_{r_2}+e^{i\theta}\ket{-L}_{r_2})\nonumber \\
	&=& \tfrac{1}{2}(_{r_2}\bra{L}+e^{-i\theta}_{\quad\:r_2}\bra{-L})(\ket{3L}_{r_2}+e^{i\theta}\ket{L}_{r_2})\nonumber \\
	&=& \tfrac{1}{2} e^{i \theta}.
	\end{eqnarray}
where in the first line we use the approximation \eqref{Phi}.

Note that the shift operator is non-Hermitian, the corresponding Hermitian operators are $\cos{2\hat{p}_{r_2}L}$ and $\sin{2\hat{p}_{r_2}L}$. In practice instead of measuring them directly, one would perform an interference experiment.

From the perspective of the external observer, we can also
understand this result, by noting that
\begin{equation} \label{rel_mom}
e^{-i 2L \hat{p}_{r_2}} = e^{i 2a \hat{p}_1} e^{-i 2b \hat{p}_2 }
\end{equation}
corresponding to a shift of particle 1 by $-2a$, and particle 2 by
$2b$ in the external reference frame. From equation (\ref{phi}) we
see that
\begin{eqnarray}
\bra{\psi}e^{i 2a \hat{p}_1  - i 2b \hat{p}_2 }\ket{\psi} &\simeq&
\bra{\psi}\left(\frac{\ket{-3a}_{1}\ket{3b}_{2}+e^{i
\theta}\ket{-a}_{1}\ket{b}_{2}}{\sqrt{2}} \right) \nonumber \\
&=&\frac{1}{2}e^{i\theta}.
\end{eqnarray}
So far everything appears perfectly consistent.

Now let us move to the case of three particles. Clearly, particle 1
can still measure the relative velocity of particle 2, even when it
has no access to particle 3. Thus particle 1 can still measure the
relative shift operator $e^{-i 2L\hat{p}_{\rr2} }$. Furthermore,
\eqref{rel_mom} still holds in the three particle case, hence using
the state \eqref{psi} as seen by the external observer, we have
\begin{equation} \label{shift_abs_3}
\bra{\Psi} e^{-i 2L \hat{p}_{r_2} } \ket{\Psi} =\bra{\Psi} e^{i
 2a \hat{p}_1} e^{-i 2b \hat{p}_2 } \ket{\Psi} =
 \frac{1}{2} e^{i\theta}.
\end{equation}
This fits with the view that particle 1 can measure the phase
without interacting with particle 3. The paradox arises when we look
at the same measurement in terms of the relative state \eqref{Psi}.
If  $e^{-i 2L \hat{p}_{r_2} }$ only acts on the relative coordinate
of particle 2, it will see only the reduced state given by
\eqref{red_state}, which has no dependence on the phase.

\subsection{Solving the paradox}

We now present the solution to the paradox. The crucial (and
surprising) observation is that $e^{-i 2L \hat{p}_{r_2} }$ actually
shifts the relative coordinate of particle 3 as well as that of
particle 2. The root of the problem is that we have been considering
relative position and momentum operators which are not canonically
conjugate to each other. When we considered the case of only two
particles $\hat{p}_{r_2}$ was the momentum canonically conjugate to
$\hat{x}_{r_2}$. However, when we move to the situation with three
particles, this is no longer the case, as $\hat{p}_{r_2}$ does
not commute with $\hat{x}_{r_3}$. Indeed it can be immediately
checked that
\begin{eqnarray}\label{q3P2} [\hat{x}_{r_3},\hat{p}_{r_2}] =  i\frac{\mu_{12}}{m_1} \neq 0. \end{eqnarray}

At this point it is important to recall that canonical observables
play a fundamental role, because they are the building blocks of the
Hilbert space and underlie its structure. Specifically, the tensor
product structure of the Hilbert space is defined through a set of
canonical observables; the eigenstates of these observables define
the natural basis. From this point of view, it now appears judicious
to identify the relative momentum operators canonically conjugate to
the relative position operators $\hat{x}_{cm}$, $\hat{x}_{r_2}$ and
$\hat{x}_{r_3}$. That is, we are looking for operators
$\hat{\pi}_{cm}, \hat{\pi}_{r_2}$, and $\hat{\pi}_{r_3}$ satisfying
the canonical commutation relations
$[\hat{x}_\alpha,\hat{\pi}_\beta]=i\delta_{\alpha\beta}$. A lengthy but straightforward
calculation leads to the following set of operators:
\begin{eqnarray}\nonumber \hat{\pi}_{cm} &=&\hat{p}_1+\hat{p}_2+\hat{p}_3,\\
\hat{\pi}_{r_2}
&=&-\frac{m_2}{M}\hat{p}_1+\left(1-\frac{m_2}{M}\right)\hat{p}_2-\frac{m_2}{M}\hat{p}_3,\\\nonumber
\hat{\pi}_{r_3}
&=&-\frac{m_3}{M}\hat{p}_1-\frac{m_3}{M}\hat{p}_2+\left(1-\frac{m_3}{M}\right)\hat{p}_3.
\end{eqnarray}

Next, let us re-express the relative shift operator $e^{-i2L
\hat{p}_{r_2} }$ in terms of these canonically conjugate momentum
operators:
\begin{eqnarray} e^{-i 2L \hat{p}_{r_2} } = e^{-i 2L \hat{\pi}_{r_2}} e^{-i 2L \frac{\mu_{12}}{m_1} \hat{\pi}_{r_3}}=e^{-i 2L \hat{\pi}_{r_2} } e^{-i 2a \hat{\pi}_{r_3} }, \end{eqnarray} where we have used the fact that $\frac{\mu_{12}}{m_1}L=a$. From the above expression, it becomes clear that $e^{-i 2L \hat{p}_{r_2} }$ acts \emph{non-locally} on the Hilbert space spanned by the relative co-ordinates $x_{r_2}$ and $x_{r_3}$, shifting both co-ordinates. Previously we argued that the inaccessibility of particle 3 by particle 1 would allow us to trace over $x_{r_3}$. However, this argument does not hold, because the relative shift operator $e^{-i2L \hat{p}_{r_2} }$ also affects $x_{r_3}$.

We can now finally understand from particle 1's perspective how it
can extract the phase from the entangled state $\eqref{Psi}$, as
\begin{eqnarray}\!\! e^{-i 2L \hat{p}_{r_2}}\ket{\Psi}\!\! &=&\!\!  e^{-i 2L \hat{\pi}_{r_2} } e^{-i 2a \hat{\pi}_{r_3}} \ket{\Psi} \\
&\simeq&\!\!
\ket{\frac{m_3c}{M}}_{cm}\!\!\left(\frac{\ket{3L}_{r_2}\ket{c+3a}_{r_3}\!+\!e^{i
\theta}\ket{L}_{r_2}\ket{c+a}_{r_3}}{\sqrt{2}}\! \right) \nonumber
\end{eqnarray}
and therefore $\bra{\Psi}e^{-i 2L \hat{p}_{r_2}}\ket{\Psi} =
\frac{1}{2} e^{i \theta}$ in agreement with \eqref{shift_abs_3}.

This solves the paradox. From particle 1's perspective, it seemed
impossible to retrieve the phase without interacting with particle
3, due to the entanglement present. However, here we see that an
operation that particle 1 can carry out on particle 2 alone
(essentially a measurement of its relative velocity),  also
indirectly affects the relative position of particle 3, allowing the
phase to be extracted even when particle 3 is physically
inaccessible.

The final piece of the puzzle is to understand physically the meaning of the apparent
non-locality of the shift operator; More precisely why it acts not only on the relative co-ordinate $x_{r_2}$ but also on $x_{r_3}$. The key observation is that particle 1 has a finite mass. Thus when it shifts particle 2 it also necessarily
shifts itself, since the centre of mass of particle 1 and 2 must remain fixed. This
shift in particle 1's position changes the relative position of
particle 3 versus particle 1. In other words, when particle 1 performs an interference experiment on particle 2, the relative distance between particles 1 and 3 is also necessarily changed. Finally, it is worth noting
that in the limit where particle 1 becomes very massive, and thus
experiences no back-action, the non-locality of the Hilbert space
disappears, i.e. the observables $\{\hat{x}_{cm},
\hat{x}_{r_2},\hat{x}_{r_3}\} $ and $\{\hat{p}_{cm},
\hat{p}_{r_2},\hat{p}_{r_3}\} $ become once again canonically conjugate, as seen
from Eq. \eqref{q3P2}.

The above illustrates that the Hilbert space has a rich structure. In the two particle case the Hilbert space possesses the normal bipartite structure that we expect, namely $\mathcal{H} = \mathcal{H}_1 \otimes \mathcal{H}_2 = \mathcal{H}_{cm}\otimes\mathcal{H}_{r_2}$, where $\mathcal{H}_{cm}$ describes the centre of mass degrees of freedom and $\mathcal{H}_{r_2}$ describes the degrees of freedom of particle 2 relative to particle 1. In particular, the eigenstates of relative position and of relative momentum belong to the relative Hilbert space, $\ket{x}_{r_2}$, $\ket{p}_{r_2} \in \mathcal{H}_{r_2}$ \footnote{Strictly speaking, position and momentum states are un-normalizable, and hence not elements of the Hilbert space. However, we can consider `approximate eigenstate' wavepackets with arbitrarily well-defined positions or momenta which belong to $\mathcal{H}_{r_2}$, and our statements can be taken to refer to these.}, and each can be written as a superposition of the other, e.g. $\ket{p}_{r_2} = \int c_p(x)\ket{x}_{r_2}dx$, where $c_p(x) \propto e^{ipx}$.

One is tempted to think that when we add a third particle and go to relative co-ordinates all we have to do is simply add on the Hilbert space for the relative degrees of freedom for this third particle, i.e. that $\mathcal{H}~=~\mathcal{H}_1 \otimes \mathcal{H}_2\otimes\mathcal{H}_3 =\mathcal{H}_{cm}\otimes\mathcal{H}_{r_2}\otimes\mathcal{H}_{r_3}$. However, this cannot be the case -- indeed from Eq. \eqref{q3P2} we already know that the operator $\hat{p}_{r_2}$ does not commute with $\hat{x}_{r_3}$, hence they cannot live in two different Hilbert spaces. The Hilbert spaces $\mathcal{H}_{r_2}$ and $\mathcal{H}_{r_3}$ therefore can have no meaning. In other words it is now no longer possible to have a Hilbert space such as $\mathcal{H}_{r_2}$ which contains both eigenstates of the relative position and momentum of particle 2 relative to particle 1; this also implies that the decomposition $\ket{p}_{r_2} = \int c_p(x)\ket{x}_{r_2}dx$ is no longer possible. In fact there exists two different decompositions of the Hilbert space, $\mathcal{H} = \mathcal{H}_{\cm}\otimes \mathcal{H}_{r_2}^x \otimes \mathcal{H}_{r_3}^x$ and $\mathcal{H} = \mathcal{H}_{\cm}\otimes\mathcal{H}_{r_2}^p \otimes \mathcal{H}_{r_3}^p$. In the first decomposition, the spaces are spanned by the eigenstates of relative position, $\hat{x}_{r_2}$ and $\hat{x}_{r_3}$ and of their conjugate momenta, $\hat{\pi}_{r_2}$ and $\hat{\pi}_{r_3}$. In the second decomposition the spaces are spanned by the eigenstates of relative momenta, $\hat{p}_{r_2}$ and $\hat{p}_{r_3}$ and their conjugate position, which we shall denote $\hat{q}_{r_2}$ and $\hat{q}_{r_3}$.

Following from the above, it is insightful to reverse the problem, that is to
start from the relative momentum operators and find explicitly their conjugate relative positions. Let us first complete the set
of relative momentum operators:
\begin{eqnarray}\nonumber \hat{p}_{cm} &=& \hat{p}_1+\hat{p}_2+\hat{p}_3  \\ \hat{p}_{r_2} &=& \mu_{12}(\frac{\hat{p}_2}{m_2}-\frac{\hat{p}_1}{m_1}) \\\nonumber \hat{p}_{r_3} &=& \mu_{13}(\frac{\hat{p}_3}{m_3}-\frac{\hat{p}_1}{m_1})
\end{eqnarray}
where $\hat{p}_{cm}$ is the total momentum, and $\hat{p}_{r_2}$ and
$\hat{p}_{r_3}$ are the relative momenta of particles 2 and 3 versus
particle 1. Now we look for the relative position operators
canonically conjugate to these relative momentum operators  (i.e.
satisfying $[\hat{q}_\alpha,\hat{p}_\beta]=i\delta_{\alpha\beta}$).
Again, a lengthy but straightforward calculation leads to
\begin{eqnarray}\nonumber \hat{q}_{cm}&=&\frac{m_1}{M}\hat{x}_1 + \frac{m_2}{M}\hat{x}_2 + \frac{m_3}{M}\hat{x}_3,\\
\hat{q}_{r_2}&=&-\frac{\mu_{13}\gamma}{m_3}\hat{x}_1 +
\gamma\hat{x}_2 -\frac{\mu_{13}\gamma}{m_1}\hat{x}_3,\\\nonumber
\hat{q}_{r_3}&=&-\frac{\mu_{12}\gamma}{m_2}\hat{x}_1
-\frac{\mu_{12}\gamma}{m_1}\hat{x}_2 + \gamma\hat{x}_3.
\end{eqnarray} with $\gamma\equiv
\frac{m_1m_2m_3}{M\mu_{12}\mu_{13}}$.

Re-writing the state \eqref{psi} in terms of these new $\hat{q}$
co-ordinates, we find that
\begin{eqnarray} \ket{\Psi} \simeq \ket{\frac{m_3c}{M}}_{cm}\!\!\left( \frac{ \ket{L- \alpha c}_{q_{r_2}}\!+ e^{i\theta}\ket{-L-\alpha c}_{q_{r_2}}}{\sqrt{2}}\!\!\right)
\!\ket{\gamma }_{q_{r_3}} \end{eqnarray} where $\alpha = \frac{\gamma
\mu_{13}}{m_1}$. Note that with these co-ordinates, the state is
fully separable. In this basis, the shift operator acts `locally' on
the $\hat{q}_{r_2}$ coordinate:
\begin{eqnarray}  e^{-i2L \hat{p}_{r_2} }\ket{\Psi}\!\! &\simeq& \!\!
\ket{\frac{m_3c}{M}}_{cm}\!\!\left(\!\frac{ \ket{3L- \alpha c}_{q_{r_2}}
\!+\! e^{i\theta}\ket{L-\alpha c}_{q_{r_2}}}{\sqrt{2}}\! \right)\!
\ket{\gamma }_{q_{r_3}} \nonumber
\end{eqnarray}
and hence $\bra{\Psi}e^{-i 2L \hat{p}_{r_2}}\ket{\Psi} = \frac{1}{2}
e^{i \theta}$ as expected

In conclusion, we have seen that the relative position and relative
momentum operators, which are the physically relevant operators from
the perspective of an internal observer, are not canonically
conjugate. This has the consequence that some operators (such as
shifting particle 2 relative to particle 1) have an apparently
`non-local' effect. Consequently, one cannot trace over the relative positions of
inaccessible particles to give a description of the remaining
accessible degrees of freedom.

Note that in the classical case, although there is no entanglement, it is still true that the usual
relative positions and momenta are not canonically conjugate.  
There are two crucial distinctions however: Firstly, a quantum measurement of  $\hat{p}_{r_2}$ must disturb $\hat{x}_{r_3}$ due to their non-commutativity, whilst classically a measurement can always be taken to produce an arbitrarily small disturbance. Secondly, the classical relative positions and momenta of particle 2 and 3 can all be independently specified, whilst quantum mechanically there is no state in which $\hat{p}_{r_2}$ and $\hat{x}_{r_3}$ are both simultaneously well defined. Hence in quantum theory, the relative state is inherently non-local, with particle 2 and particle 3  intertwined.

\section{Absolute frames of reference?}

In all the cases we have considered, we have  begun by defining the state in an external  frame, and then moved to relative co-ordinates. When we refer to the centre of mass coordinate, we are really describing its position in this external frame. But does our analysis depend on which external frame we initially choose? Do we require an external  frame of reference at all? If so, what properties should it have? These issues are of particular importance in section II,  where the experimental results depend critically on whether the relative coordinate is entangled with the centre of mass coordinate or not.

The first thing to note is that if the centre of mass is in a superposition of different places in relation to one frame of reference it is also in a superposition relative to any other frame of reference that is classically related to the first frame, i.e. which has a well defined displacement and speed relative to this frame -- a Galilean transformation. So certainly we do not single out one particular frame amongst all the frames related by a Galilean transformation.

However, in quantum mechanics, one might consider alternative `non-classical' transformations of the external reference frame such as moving to a frame in a superposition of different locations relative to the original frame \cite{bartlett09,BarRudSpe06,KitMayPre04}. Looking at figure (1b) it is possible to imagine a frame of reference in which the interferometer is not in a superposition but located at the origin -- just go to a frame of reference which is in a superposition versus the original one and correlated with the location of the interferometer. In this frame of reference, the position of the centre of mass will be in the same place in each branch of the superposition, and hence it will factorise. Does it mean that we are now able to measure interference? Of course, whether or not we are able to see interference is just a question of internal measurements and should be independent of the external reference frame, so there is an inconsistency. Does this therefore mean that such reference frames which are in a superposition are not permitted? Is it the case that there exists an absolute \emph{set} of reference frames which are related by Galilean transformations but all other frames which are in superposition relative to them do not exist? We seem to have run into a serious problem.

These problems have several aspects, and below we will address them in turn.

The first issue is whether we can determine the existence or properties of the external frame from inside the system alone. i.e. without interacting directly with the external frame. Unsurprisingly, the answer to this question is no. As in classical mechanics, we are at liberty to describe any physical situation only in terms of relative co-ordinates. For each of the physical situations we consider, the correct relative description is obtained by tracing out the centre of mass degree of freedom, yielding a state of only  the relative degrees of freedom. The Hamiltonian can also always be decomposed into two commuting terms, one of which is a function of the total momentum alone, and the other is a function of only the relative positions and momenta. Keeping only the second term will correctly describe the evolution of all the relative degrees of freedom without reference to the centre of mass.

However, our results  highlight two curious aspects of this approach: (i) That some situations which can be described by pure states in an external frame must be described using mixed states in terms of relative co-ordinates \cite{bartlett09}. (ii) That the concept of locality is much harder to pin down in terms of relative co-ordinates, as the relative position and momentum of one particle do not commute with those of the other particles. Hence we cannot simply trace out the relative co-ordinates of particles we are not interested in, and operations by the observer on one particle will affect the relative co-ordinates of the others.

Another interesting consideration is that of transforming between different external reference frames, rather than eliminating the external frame entirely.

There are two ways in which we can think of an external reference frame, either (i) as an abstract labelling of space-time, or (ii) as a physical laboratory containing rulers and clocks. In case (i) transforming between reference frames is a purely mathematical transformation, a simple relabelling of the co-ordinates. In case (ii) the issue of transformation is more subtle. There are actually two subcases. The first way (iia) is to consider one single physical frame and going to a different frame being a purely mathematical transformation on the co-ordinates indicated on this physical frame. For example, the physical frame may be a ruler, and we can go to a displaced frame by simply relabelling the marks of the ruler. The second way (iib) is to have two actual physical systems. For example, we could take two rulers, one displaced relative to the other.

Classically these different approaches are completely equivalent to one another. Quantum Mechanically however, the situation is more complicated. Let us illustrate this in the example of section II, where we analysed a particle's interaction with an interferometer in two different situations. In the second case (depicted in Fig. (1b)), the interferometer was prepared in a superposition of locations in the external reference frame, whilst the particle had a well-defined location.  Consider now moving to a new reference frame in which the interferometer is localised near the origin.

Let us see what happens if we attempt to change co-ordinates via a simple relabellings as in (i) or (iia). The description of the state in the new reference frame cannot be given by a pure state, because the only pure state in which the particle can be found with equal probability at $L$ and $-L$ and the interferometer always found at the origin is of the form \eqref{eqn:setup_a} up to some well defined relative phase between $\ket{-L}_{p}$ and $\ket{L}_{p}$. This state however leads to different physical predictions than those made in the original frame, namely that there is interference in the interferometer. Alternatively, we might seek to describe the state in the new reference frame using a mixed state, but then the transformation between reference frames must be irreversible, and we would not be able to transform back to the first reference frame and recover the original description. All of these problems stem from the fact that it makes no sense to relabel in superposition. Superposition and entanglement are properties of the states of physical objects -- it is the physical object whose states can be in superposition or entangled; there is no notion of abstract mathematical relabellings being in superposition.

Given the above difficulties, we now turn to the case (iib), in which different \emph{external} frames are considered as distinct physical laboratories. We take these laboratories to be non-interacting rigid quantum objects of very large mass, such that the positions and momenta of objects in the system relative to the laboratory obey essentially the usual canonical commutation relations (consider (\ref{q3P2}) in the limit as $m_1 \rightarrow \infty$).

We can now approach the paradox discussed at the beginning of this section. Consider the situation depicted in figure (1b) but now modified to include two physical frames, the original frame $F_1$ and a second frame $F_2$, in which the interferometer is located at the origin. In order for the interferometer to be located at the origin in $F_2$ it must be the case that it is actually entangled with the position of the interferometer. As we will show below, given that we now consider explicitly two frames of reference, the predictions made from each of them are in fact consistent, and there is no longer a paradox.

As seen from frame $F_1$, the state of the particle, interferometer and $F_2$ is
\begin{equation}
\ket{\psi'_b} =
\frac{1}{\sqrt{2}}  \left(\ket{-L}_{i} \ket{-L}_{F_2}+\ket{L}_{i} \ket{L}_{F_2} \right)\ket{0}_{p}.
\end{equation}
Note that this state is now different from (\ref{eqn:setup_b}), which did not include the frame $F_2$. Rewriting this in terms of the relative co-ordinates $r_p$ and $r_i$ of the particle and  interferometer relative to the $F_2$, we obtain
\begin{equation}
\ket{\psi'_b} \simeq
\frac{1}{\sqrt{2}} \left( \ket{-L}_{cm} \ket{L}_{r_p}  + \ket{L}_{cm} \ket{-L}_{r_p} \right) \ket{0}_{r_i}.
\end{equation}
Tracing out the centre of mass, we see that the description of the state in $F_2$ is the mixed state $\rho= \frac{1}{2}(\proj{L} + \proj{-L})_{r_p} \otimes \proj{0}_i$, which will not yield interference. Hence in this case the description relative to the original external frame $F_1$ and the new frame $F_2$ both give the same physical predictions.

Let us now modify the set-up in figure (1a) in a similar manner, introducing two frames, the original frame $F_1$ and a new frame $F_2$ in which the particle is located at the origin. This modified set-up is now completely equivalent to the modified set-up of figure (1b) with the role of the two frames interchanged. As such there cannot be interference in the modified set-up (1a). This may seem surprising, since in the original set-up there was interference.

Indeed, the state now, instead of being (\ref{eqn:setup_a}) is given by
\begin{equation} \label{eqn:lab}
\ket{\psi'_a} =
\frac{1}{\sqrt{2}} \left( \ket{L}_{p} \ket{L}_{F_2} +  \ket{-L}_{p} \ket{-L}_{F_2} \right) \ket{0}_i,
\end{equation}
as seen from $F_1$. If the frame $F_2$ is much more massive than the interferometer, this state can be expressed in relative co-ordinates as
\begin{equation}
\ket{\psi'_a}\simeq
\frac{1}{\sqrt{2}} \left( \ket{L}_{cm} \ket{-L}_{r_i}  + \ket{-L}_{cm} \ket{L}_{r_i} \right) \ket{0}_{r_p}.
\end{equation}
By tracing out the centre of mass, we  obtain a mixed state from the perspective of the interferometer, hence interference will be lost \footnote{In fact, we can see immediately this by tracing out the lab in (\ref{eqn:lab}), as the position of the lab acts as a `which-way' detector for the photon.}. Introducing the second physical frame $F_2$ into the state has therefore \emph{changed} the physical predictions (destroying the interference). This may seem surprising, since all the measurements we perform involve only the particle and the interferometer and we do not touch the frame $F_2$ during the experiment. Why does then the mere existence of the frame $F_2$ change the physics?

This stems from the fact that we wanted to introduce the frame $F_2$ in such a way that it is entangled with our system. In the classical case we can introduce physical frames at will without any need to interact with the system and hence without changing its behaviour. However, in order to prepare an entangled state, there must be an interaction between the system and the frame. This is when the system is affected.

Finally, we can consider introducing a quantum frame that is not entangled with the system, but is in a superposition of locations. After tracing out the centre of mass, the situation observed from the perspective of such a frame will be the same as if it were prepared in a mixture of locations. This will not change the physical predictions relating to internal properties of the system (such as interference in our interferometry experiments), but it will nevertheless constitute an irreversible transformation of the original state of the system, mapping pure states to mixed states and losing information.

\section{Conclusion}
We investigated the physics of quantum reference frames by studying simple situations, involving only a few particles, whereby we promoted one of these particles to the role of a quantum observer. We then asked the question of how this quantum observer would describe the rest of the system. We first considered the case in which one of the particles is in superposition. We argued that the centre of mass, in particular whether it is entangled to the relative co-ordinates or not, is a central issue. Second, we considered the situation in which two particles of the system are entangled such that the centre of mass and relative position are unentangled. We then introduced a third particle which led us to realise that the Hilbert space of the quantum observer is inherently non-local. Finally we discussed the issue of an absolute quantum frame of reference and showed that in quantum theory, introducing  external reference frames can affect the behaviour of the system being studied in particular, when the additional reference frame and the system are taken to be entangled.

In the future it would be interesting to explore further the consequences of the non-local structure of the Hilbert space of quantum observers. Furthermore it would be important to go beyond the one-dimensional case as new physical quantities arise in this setting, for example in two-dimensions and above we must also consider angular momentum.

\section*{Acknowledgements}
We Acknowledge financial support from CNPq/Brazil, the UK EPSRC, the projects QIP-IRC and QAP. AJS acknowledges support from the Royal Society.

\appendix
\section{Finite width packet effects}

In this appendix, we further discuss the subtle issues of entanglement due to finite wavepacket width that were introduced in section \ref{sec:2_particle}. We note in the main text that a product state of two wavepackets in the external frame of reference will generally become entangled when expressed in terms of relative co-ordinates. For clarity, we have made an approximation in which this entanglement of microscopic fluctuations is ignored, such that a product state in the external reference frame remains a product state when transformed into relative co-ordinates. We therefore write $\ket{a}_1 \ket{b}_2 \simeq \ket{\alpha}_{cm} \ket{\beta}_r$ throughout.

Note that although fluctuations in the centre of mass and relative coordinate only extend over a small physical distance, this sense of approximation does not correspond to equivalence between the states for all measurements  (as the trace distance would). In particular,  fine-grained measurements on the scale of a wavepacket might be able to distinguish $\ket{a}_1 \ket{b}_2$ from any particular product state $\ket{\alpha}_{cm} \ket{\beta}_r$. However, such measurements are not relevant to our arguments.

Furthermore, for some special choices of initial wavepacket in the two particle case, our approximation holds exactly. For example consider a product state of two Gaussian wavepackets in the external frame:
\begin{equation}
	\psi(x_1-a)\phi(x_2-b) = e^{-\frac{1}{2\Delta_1^2}(x_1-a)^2}e^{-\frac{1}{2\Delta_2^2}(x_2-b)^2}
\end{equation}
i.e. particle 1 is centred around the position $a$ and has uncertainty $\Delta_1$ and particle 2 is centred around $b$ with uncertainty $\Delta_2$. Moving to relative co-ordinates, using equations (\ref{eqn:x_cm})-(\ref{eqn:x_r}) from the main text, we find the state transforms into
	\begin{eqnarray}
		\Psi(x_\cm,x_\rr) &=& e^{-\frac{1}{2\Delta_1^2}(x_\cm-\frac{m_2}{M}x_\rr-a)^2}e^{-\frac{1}{2\Delta_2^2}(x_\cm+\frac{m_1}{M}x_\rr-b)^2}  \\
		&=& e^{-\frac{1}{2\Delta_c^2}(x_\cm - \alpha)^2}e^{-\frac{1}{2\Delta_r^2}(x_\rr-\beta)^2}e^{2 \gamma (x_\cm - \alpha) (x_\rr - \beta)} \nonumber
	\end{eqnarray}
	where
	\begin{eqnarray}
		\Delta_c^2 &=&\Delta_1^2\Delta_2^2/(\Delta_1^2+\Delta_2^2)  \\
		\Delta_r^2 &=& M^2\Delta_1^2\Delta_2^2/(m_1^2\Delta_1^2+m_2^2\Delta_2^2)   \\
		\alpha &=& (m_1 a + m_2 b)/M \\
		\beta &=& b - a \\
		\gamma &=& (m_2\Delta_2^2 - m_1\Delta_1^2)/M\Delta_1^2\Delta_2^2
	\end{eqnarray}
Hence in general, the relative and centre of mass coordinate will be entangled. However, if we choose $m_1\Delta_1^2 = m_2\Delta_2^2$ we obtain $\gamma =0$ and recover a product state. By taking the initial wavepackets to be Gaussians satisfying $m_1\Delta_1^2 = m_2\Delta_2^2$ and the transformed wavepackets to be Gaussians of widths $\Delta x_c$ and $\Delta x_r$, we could therefore eliminate our approximation and have $\ket{a}_1 \ket{b}_2 = \ket{\alpha}_{cm} \ket{\beta}_r$. However, we emphasise that our central observations extend beyond this special case. All we really require are that wavepackets are sufficiently small, and that all wavepackets for the same particle in an initial  state have the same form (i.e. they are translated copies of each other).

With three particles, the centre of mass will again decouple from the relative co-ordinates when the initial wavepackets are Gaussians satisfying $m_1\Delta_1^2 = m_2\Delta_2^2 = m_3\Delta_3^2$. However, in this case there will remain some residual entanglement  between the relative co-ordinates $x_{r_2}$ and $x_{r_3}$ in the transformed state. If we wished to include this entanglement explicitly, we could rewrite (\ref{Psi}) as
\begin{equation}
\ket{\Psi}= \left|\frac{m_3}{M}c\right\rangle_{cm}\left(\frac{\ket{L, c+a}_{r_2 r_3} +e^{i
\theta}\ket{-L, c-a}_{r_2 r_3}}{\sqrt{2}} \right).
\end{equation}
where $\ket{a,b}_{r_2 r_3}$ represents an entangled wavepacket in the relative co-ordinates, centred on $r_2=a$ and $r_3 =b$. The reduced state of $r_2$ given by (\ref{red_state}) would then be modified to
\begin{equation}
\rho_{r_2} = \tr_{cm, r_3} (\ket{\Psi}\bra{\Psi}) \approx
\frac{ \rho_{+L} + \rho_{-L}}{2},
\end{equation}
where $\rho_{\pm L} = \tr_{r_3} (\proj{\pm L, c \pm a})$ are mixed wavepackets centred on $r_2 = \pm L$, and we have used the fact that $ \tr_{r_3} (\op{L, c+a}{-L, c-a}) \approx 0$. Note that the phase is still unobservable in this state, and hence the conclusions are identical to those in the approximated case.

Similarly, a detailed investigation of our other results without approximations or a particular choice of initial wavepackets yields the same conclusions throughout, but with more complexity.

\end{document}